\documentclass[letter]{aa} 
\usepackage{graphicx}
\usepackage{txfonts}
\usepackage{natbib}
\bibpunct{(}{)}{;}{a}{}{,}
\begin{document}
   \title{Erratum: SPITZER-IRS spectral fitting of discs around binary post-AGB stars.
}

   \author{
          C. Gielen\inst{1}\fnmsep \thanks{Postdoctoral Fellow of
the Fund for Scientific Research, Flanders}
          \and
          H. Van Winckel\inst{1}
          \and
          M. Min \inst{17}
          \and
          L.B.F.M. Waters \inst{1,2}
          \and
          T. Lloyd Evans \inst{3}
          \and
          M. Matsuura \inst{4,5}
          \and
          P. Deroo \inst{6}
          \and
          C. Dominik \inst{2,7}
          \and
          M. Reyniers\inst{8}
          \and
          A. Zijlstra\inst{9}
          \and
          K.~D. Gordon\inst{11}
          \and
          F. Kemper\inst{9}
          \and
          R. Indebetouw\inst{12,16}
          \and
          M. Marengo\inst{13}
          \and
          M. Meixner\inst{11}
          \and
          G.C. Sloan\inst{14}
          \and
          A.~G.~G.~M. Tielens\inst{15}
          \and
          P.~M. Woods\inst{9}
          }


   \institute{Instituut voor Sterrenkunde,
              Katholieke Universiteit Leuven, Celestijnenlaan 200D, 3001 Leuven, Belgium\\ 
	      \email{clio.gielen@ster.kuleuven.be}
              \and
              Sterrenkundig Instituut 'Anton Pannekoek', 
              Universiteit Amsterdam, Kruislaan 403, 1098 Amsterdam, The Netherlands
              \and
              SUPA, School of Physics and Astronomy, 
              University of St Andrews, North Haugh, St Andrews, Fife KY16 9SS, United Kingdom 
              \and
              UCL-Institute of Origins, Department of Physics and Astronomy, University College London, Gower Street, London WC1E 6BT, United Kingdom 
              \and
              UCL-Institute of Origins, Mullard Space Science Laboratory, University College London, Holmbury St. Mary, Dorking, Surrey RH5 6NT, United Kingdom 
              \and
              Jet Propulsion Laboratory, 4800 Oak Grove Drive, Pasadena, CA 91109, US  
              \and
              Department of Astrophysics, Radboud University Nijmegen, PO Box 9010, 6500 GL Nijmegen, The Netherlands 
              \and
              The Royal Meteorological Institute of Belgium, Department Observations, 
              Ringlaan 3, 1180 Brussels, Belgium 
              \and
              Jodrell Bank Centre for Astrophysics, Alan Turing Building, University of Manchester, 
              Oxford Road, Manchester, M13 9PL, United Kingdom
              \and
              SUPA, School of Physics and Astronomy, 
              University of St Andrews, North Haugh, St Andrews, Fife KY16 9SS, United Kingdom 
              \and
              Space Telescope Science Institute, 3700 San Martin Drive, Baltimore, MD 21218, USA 
              \and
              Department of Astronomy, University of Virginia, PO Box 3818, Charlottesville, VA 22903-0818, USA 
              \and
              Harvard-Smithsonian Center for Astrophysics, 60 Garden Street, MS 65, Cambridge, MA 02138-1516, USA 
              \and
              Department of Astronomy, Cornell University, Ithaca, NY 14853-6801, USA 
              \and
              Leiden Observatory, J.H. Oort Building, Niels Bohrweg 2, 2333 CA Leiden, The Netherlands 
              \and
              National Radio Astronomy Observatory, 520 Edgemont Road, Charlottesville, VA 22906, USA 
              \and
	      Astronomisch Instituut Utrecht, Universiteit Utrecht, Princetonplein 5, 3584 CC Utrecht, The Netherlands 
}

   \date{Received ; accepted }

  \abstract
   {}
   {no abstract}
   {no abstract}
   {no abstract}
   {}
   \maketitle
%

Recently, we have discovered an error in our Monte-Carlo spectral fitting routine, more specifically where
the errors on the fluxes were rescaled to get a reduced $\chi^2$ of $1$. The rescaled errors were
too big, resulting in too wide a range of 'good' fits in our 100 step Monte-Carlo routine.

This problem affects Figs.\,7-9 and Table\,A.1-A.2 in \citet{gielen08}, Table\,3 in \citet{gielen09},
and Table\,4 in \citet{gielen09b}.

We corrected for this error and present the new values and errors in the tables below.
The new values and errors nearly all fall within the old error range. Our best $\chi^2$
values and overall former scientific results are not affected.
With these new errors some possible new trends in the dust parameters might be observed.
These will be discussed in an upcoming paper where we extend the sample presented in 
\citet{gielen08} with newly obtained SPITZER-IRS data. 

\begin{figure}
\vspace{4cm}
\hspace{0cm}
\resizebox{9cm}{!}{ \includegraphics{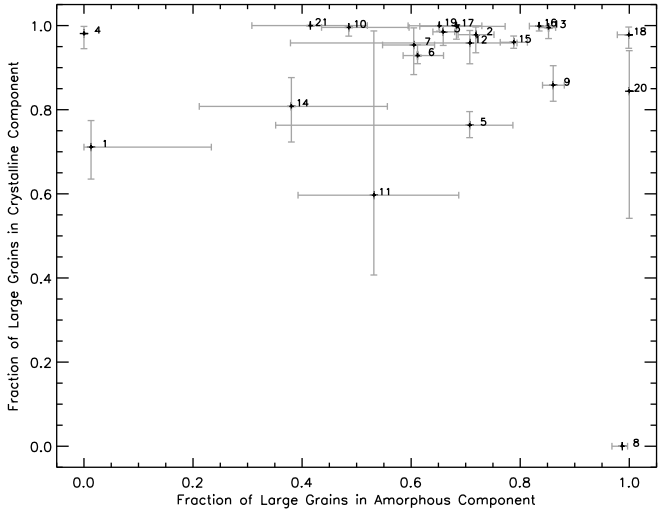}}
\caption{\textit{Erratum for Fig.\,7 in \citet{gielen08}:} The fraction of large grains in the amorphous component versus the fraction of large
grains in the crystalline component, using the fitting with grain sizes of 0.1\,$\mu$m and 2.0\,$\mu$m. 
Crystalline grains are almost completely made up of large 2.0\,$\mu$m grains.}
\end{figure}

\begin{figure}
\vspace{0cm}
\hspace{0cm}
\resizebox{9cm}{!}{ \includegraphics{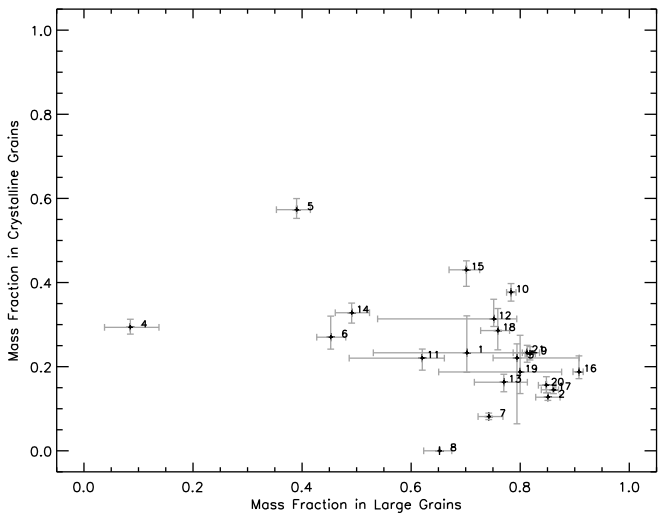}}
\caption{\textit{Erratum for Fig.\,8 in \citet{gielen08}:} The mass fraction in large grains (4.0\,$\mu$m) plotted against the mass fraction in crystalline grains, as derived
from our best-fit parameters.}
\end{figure}

\begin{figure}
\vspace{4cm}
\hspace{0cm}
\resizebox{9cm}{!}{ \includegraphics{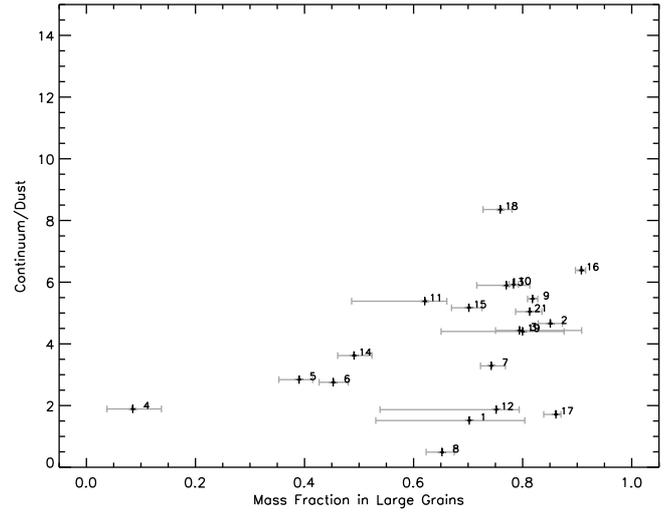}}
\caption{\textit{Erratum for Fig.\,9 in \citet{gielen08}:} The continuum-to-dust ratio of the observed spectra plotted against the mass fraction on large grains (4.0\,$\mu$m).}
\end{figure}

\onecolumn
\begin{table}[h]
\caption{\textit{Erratum for Table\,A.1 in \citet{gielen08}}: Best-fit parameters deduced from our full spectral fitting.
Listed are the $\chi^2$, dust and continuum temperatures and their relative fractions.}
\centering
\begin{tabular}{llrllllll}
\hline \hline
N$^\circ$ & Name & $\chi^2$ &  $T_{dust1}$ & $T_{dust2}$ & Fraction & $T_{cont1}$ & $T_{cont2}$ & Fraction \\
       &      &          &     (K)     & (K)         & $T_{dust1}$- $T_{dust2}$    & (K)         & (K)         & $T_{cont1}
$-$T_{cont2}$   \\
\hline
  1 &EP\,Lyr    &56.7 &$ 100_{   50}^{   50}$ &$ 200_{   50}^{   50}$ &$ 0.90_{ 0.10}^{ 0.05}- 0.10_{ 0.05}^{ 0.10}$
&$ 200_{   50}^{   50}$ &$ 994_{ 103}^{   50}$ &$ 0.98_{ 0.01}^{ 0.01}- 0.02_{ 0.01}^{ 0.01}$\\
  2 &HD\,131356 & 3.5 &$ 200_{   50}^{   50}$ &$1000_{   50}^{   50}$ &$ 0.90_{ 0.05}^{ 0.05}- 0.10_{ 0.05}^{ 0.05}$
&$ 200_{   50}^{   50}$ &$ 500_{   50}^{   50}$ &$ 0.90_{ 0.01}^{ 0.01}- 0.10_{ 0.01}^{ 0.01}$\\
  3 &HD\,213985 & 4.4 &$ 184_{  87}^{  50}$ &$1000_{   50}^{   50}$ &$ 0.90_{ 0.05}^{ 0.05}- 0.10_{ 0.05}^{ 0.05}$ &$
 200_{   50}^{   50}$ &$ 884_{  87}^{  50}$ &$ 0.98_{ 0.01}^{ 0.01}- 0.02_{ 0.01}^{ 0.01}$\\
  4 &HD\,52961  &72.2 &$ 200_{   50}^{   50}$ &$ 800_{   50}^{   50}$ &$ 0.90_{ 0.05}^{ 0.05}- 0.10_{ 0.05}^{ 0.05}$
&$ 100_{   50}^{   50}$ &$1000_{   50}^{   50}$ &$ 0.99_{ 0.01}^{ 0.01}- 0.01_{ 0.01}^{ 0.01}$\\
  5 &IRAS\,05208        & 4.5 &$ 292_{  95}^{  50}$ &$ 923_{ 113}^{  78}$ &$ 0.80_{ 0.10}^{ 0.05}- 0.20_{ 0.05}^{ 0.1
0}$ &$ 200_{   50}^{   50}$ &$ 400_{   50}^{   50}$ &$ 0.85_{ 0.01}^{ 0.01}- 0.15_{ 0.01}^{ 0.01}$\\
  6 &IRAS\,09060        & 3.6 &$ 200_{  50}^{   50}$ &$ 728_{ 130}^{  73}$ &$ 0.90_{ 0.05}^{ 0.05}- 0.10_{ 0.05}^{ 0.
05}$ &$ 228_{ 130}^{  73}$ &$ 834_{ 237}^{ 141}$ &$ 0.93_{ 0.02}^{ 0.02}- 0.07_{ 0.02}^{ 0.02}$\\
  7 &IRAS\,09144        & 6.1 &$ 200_{   50}^{   50}$ &$ 504_{   50}^{ 111}$ &$ 0.90_{ 0.05}^{ 0.05}- 0.10_{ 0.05}^{
0.05}$ &$ 200_{   50}^{  50}$ &$ 796_{ 111}^{   50}$ &$ 0.94_{ 0.01}^{ 0.01}- 0.06_{ 0.01}^{ 0.01}$\\
  8 &IRAS\,10174        &13.9 &$ 300_{   50}^{   50}$ &$ 400_{   50}^{   50}$ &$ 0.90_{ 0.05}^{ 0.05}- 0.10_{ 0.05}^{
 0.05}$ &$ 100_{   50}^{   50}$ &$ 300_{   50}^{   50}$ &$ 0.97_{ 0.01}^{ 0.01}- 0.03_{ 0.01}^{ 0.01}$\\
  9 &IRAS\,16230        & 4.9 &$ 200_{   50}^{   50}$ &$ 500_{   50}^{   50}$ &$ 0.90_{ 0.05}^{ 0.05}- 0.10_{ 0.05}^{
 0.05}$ &$ 100_{   50}^{   50}$ &$ 500_{   50}^{   50}$ &$ 0.95_{ 0.05}^{ 0.05}- 0.05_{ 0.05}^{ 0.05}$\\
 10 &IRAS\,17038        & 2.9 &$ 317_{  50}^{  85}$ &$ 871_{  80}^{  61}$ &$ 0.80_{ 0.10}^{ 0.10}- 0.20_{ 0.10}^{ 0.1
0}$ &$ 200_{   50}^{   50}$ &$ 591_{  96}^{   50}$ &$ 0.97_{ 0.02}^{ 0.01}- 0.03_{ 0.01}^{ 0.02}$\\
 11 &IRAS\,17243        & 2.3 &$ 200_{   50}^{   50}$ &$ 486_{  89}^{  50}$ &$ 0.90_{ 0.10}^{ 0.05}- 0.10_{ 0.05}^{ 0
.10}$ &$ 200_{   50}^{   50}$ &$ 600_{   50}^{   50}$ &$ 0.90_{ 0.01}^{ 0.01}- 0.10_{ 0.01}^{ 0.01}$\\
 12 &IRAS\,19125        & 3.9 &$ 100_{   50}^{   50}$ &$ 200_{   50}^{   50}$ &$ 0.90_{ 0.05}^{ 0.05}- 0.10_{ 0.05}^{
 0.05}$ &$ 482_{ 309}^{  50}$ &$ 788_{ 206}^{  50}$ &$ 0.86_{ 0.05}^{ 0.01}- 0.14_{ 0.01}^{ 0.05}$\\
 13 &IRAS\,19157        & 5.5 &$ 200_{   50}^{   50}$ &$ 695_{ 106}^{   50}$ &$ 0.90_{ 0.05}^{ 0.05}- 0.10_{ 0.05}^{
0.05}$ &$ 200_{   50}^{   50}$ &$ 705_{   50}^{ 106}$ &$ 0.96_{ 0.01}^{ 0.01}- 0.04_{ 0.01}^{ 0.01}$\\
 14 &IRAS\,20056        & 3.8 &$ 100_{   50}^{   50}$ &$ 200_{   50}^{   50}$ &$ 0.10_{ 0.10}^{ 0.20}- 0.90_{ 0.20}^{
 0.10}$ &$ 200_{   50}^{   50}$ &$ 600_{   50}^{   50}$ &$ 0.88_{ 0.01}^{ 0.01}- 0.12_{ 0.01}^{ 0.01}$\\
 15 &RU\,Cen    & 3.4 &$ 287_{  91}^{  50}$ &$ 575_{ 176}^{  50}$ &$ 0.90_{ 0.30}^{ 0.05}- 0.10_{ 0.05}^{ 0.30}$ &$ 2
00_{   50}^{   50}$ &$ 599_{   50}^{   50}$ &$ 0.99_{ 0.01}^{ 0.01}- 0.01_{ 0.01}^{ 0.01}$\\
 16 &SAO\,173329        & 3.1 &$ 200_{   50}^{   50}$ &$ 702_{   50}^{ 139}$ &$ 0.90_{ 0.05}^{ 0.05}- 0.10_{ 0.05}^{
0.05}$ &$ 200_{   50}^{   50}$ &$ 600_{   50}^{   50}$ &$ 0.93_{ 0.01}^{ 0.01}- 0.07_{ 0.01}^{ 0.01}$\\
 17 &ST\,Pup    & 8.4 &$ 200_{   50}^{   50}$ &$ 500_{   50}^{   50}$ &$ 0.90_{ 0.05}^{ 0.05}- 0.10_{ 0.05}^{ 0.05}$
&$ 200_{   50}^{   50}$ &$ 500_{   50}^{   50}$ &$ 0.95_{ 0.01}^{ 0.01}- 0.05_{ 0.01}^{ 0.01}$\\
 18 &SU\,Gem    & 1.8 &$ 154_{  54}^{ 105}$ &$ 558_{  59}^{  96}$ &$ 0.80_{ 0.10}^{ 0.10}- 0.20_{ 0.10}^{ 0.10}$ &$ 2
00_{   50}^{   50}$ &$ 800_{   50}^{  50}$ &$ 0.95_{ 0.01}^{ 0.01}- 0.05_{ 0.01}^{ 0.01}$\\
 19 &SX\,Cen    & 4.3 &$ 257_{  58}^{  50}$ &$ 968_{  69}^{  50}$ &$ 0.80_{ 0.10}^{ 0.10}- 0.20_{ 0.10}^{ 0.10}$ &$ 2
00_{   50}^{   50}$ &$ 691_{  96}^{   50}$ &$ 0.94_{ 0.01}^{ 0.01}- 0.06_{ 0.01}^{ 0.01}$\\
 20 &TW\,Cam    & 2.3 &$ 261_{  62}^{  50}$ &$ 400_{   50}^{   50}$ &$ 0.60_{ 0.05}^{ 0.10}- 0.40_{ 0.10}^{ 0.05}$ &$
 100_{   50}^{   50}$ &$ 500_{   50}^{   50}$ &$ 0.95_{ 0.01}^{ 0.01}- 0.05_{ 0.01}^{ 0.01}$\\
 21 &UY\,CMa    & 2.9 &$ 200_{   50}^{   50}$ &$ 726_{  50}^{  76}$ &$ 0.90_{ 0.05}^{ 0.05}- 0.10_{ 0.05}^{ 0.05}$ &$
 200_{   50}^{   50}$ &$ 500_{   50}^{   50}$ &$ 0.83_{ 0.01}^{ 0.01}- 0.17_{ 0.01}^{ 0.01}$\\

\hline
\end{tabular}
\end{table}

\begin{table}[h]
\caption{\textit{Erratum for Table\,A.2 in \citet{gielen08}}: Best-fit parameters deduced from our full spectral fitting. The abundances of small (2.0\,$\mu$m) and large (4.0\,$\mu$m) grains of the various
dust species are given as fractions of the total mass, excluding the dust responsible for the continuum emission.
The last column gives the continuum flux contribution, listed as a percentage of the total integrated flux over the 
full wavelength range.}
\centering
\begin{tabular}{llccccc}
\hline \hline
 N$^\circ$ & Name & Olivine & Pyroxene & Forsterite & Enstatite & Continuum\\
           &      & Small  -  Large & Small  -  Large &  Small  -   Large & Small  -   Large &\\
\hline
  1    &EP\,Lyr    &$ 0.00_{ 0.00}^{ 0.00}    -   0.00_{ 0.00}^{ 0.00}$    &$15.36_{ 9.25}^{12.96}    -  61.34_{23.01}^{10.
71}$    &$14.43_{ 2.17}^{ 4.61}    -   0.02_{ 0.02}^{ 0.00}$    &$ 0.00_{ 0.00}^{ 0.00}    -   8.85_{ 3.90}^{ 4.49}$    &$5
3.51_{ 1.43}^{ 3.36}$\\
  2    &HD\,131356    &$ 0.00_{ 0.00}^{ 0.00}    -  30.48_{ 1.44}^{ 1.50}$    &$ 2.87_{ 1.77}^{ 1.79}    -  53.91_{ 2.72}^{
 2.57}$    &$12.06_{ 0.62}^{ 0.65}    -   0.03_{ 0.03}^{ 1.00}$    &$ 0.00_{ 0.00}^{ 0.00}    -   0.65_{ 0.56}^{ 1.34}$
&$76.34_{ 0.24}^{ 0.30}$\\
  3    &HD\,213985    &$ 0.00_{ 0.00}^{ 0.00}    -  36.08_{ 2.21}^{ 2.20}$    &$12.58_{ 9.92}^{ 3.42}    -  29.27_{ 6.99}^{
25.84}$    &$ 8.02_{ 1.63}^{ 1.79}    -   7.74_{ 4.86}^{ 3.26}$    &$ 0.00_{ 0.00}^{ 0.00}    -   6.31_{ 4.45}^{ 2.09}$
&$76.21_{ 1.51}^{ 0.57}$\\
  4    &HD\,52961    &$ 0.00_{ 0.00}^{ 0.00}    -   0.00_{ 0.00}^{ 0.00}$    &$70.65_{ 1.93}^{ 1.61}    -   0.00_{ 0.00}^{
0.00}$    &$20.84_{ 3.57}^{ 3.81}    -   8.40_{ 4.94}^{ 5.18}$    &$ 0.00_{ 0.00}^{ 0.00}    -   0.12_{ 0.12}^{ 2.07}$    &
$65.66_{ 0.62}^{ 0.57}$\\
  5    &IRAS\,05208    &$ 0.00_{ 0.00}^{ 0.00}    -   9.52_{ 2.52}^{ 2.49}$    &$33.16_{ 1.29}^{ 1.52}    -   0.00_{ 0.00}^
{ 0.00}$    &$25.76_{ 1.10}^{ 1.34}    -   0.00_{ 0.00}^{ 0.00}$    &$ 2.10_{ 1.67}^{ 2.13}    -  29.47_{ 3.27}^{ 2.80}$
 &$69.06_{ 0.38}^{ 0.38}$\\
  6    &IRAS\,09060    &$ 0.06_{ 0.06}^{ 1.80}    -  32.70_{ 4.58}^{ 4.64}$    &$39.74_{ 2.66}^{ 3.00}    -   0.48_{ 0.49}^
{ 4.21}$    &$14.95_{ 1.81}^{ 1.37}    -   1.11_{ 1.11}^{ 2.97}$    &$ 0.01_{ 0.01}^{ 0.40}    -  10.95_{ 2.51}^{ 3.37}$
 &$72.43_{ 2.20}^{ 1.24}$\\
  7    &IRAS\,09144    &$ 0.00_{ 0.00}^{ 0.00}    -  39.21_{ 4.59}^{ 2.04}$    &$17.77_{ 2.65}^{ 1.54}    -  34.88_{ 3.03}^
{ 7.04}$    &$ 7.99_{ 0.63}^{ 0.75}    -   0.00_{ 0.00}^{ 0.00}$    &$ 0.00_{ 0.00}^{ 0.00}    -   0.14_{ 0.14}^{ 1.21}$
 &$72.51_{ 0.30}^{ 0.39}$\\
  8    &IRAS\,10174    &$ 8.70_{ 3.38}^{ 4.93}    -  39.99_{ 6.00}^{ 4.24}$    &$26.10_{ 3.10}^{ 2.29}    -  25.21_{ 3.46}^
{ 4.19}$    &$ 0.00_{ 0.00}^{ 0.00}    -   0.00_{ 0.00}^{ 0.00}$    &$ 0.00_{ 0.00}^{ 0.00}    -   0.00_{ 0.00}^{ 0.00}$
 &$31.48_{ 0.49}^{ 0.54}$\\
  9    &IRAS\,16230    &$ 0.00_{ 0.00}^{ 0.00}    -  47.54_{ 2.30}^{ 2.35}$    &$ 0.00_{ 0.00}^{ 0.00}    -  29.46_{ 2.39}^
{ 2.01}$    &$18.20_{ 0.98}^{ 0.90}    -   4.23_{ 1.50}^{ 1.38}$    &$ 0.00_{ 0.00}^{ 0.00}    -   0.58_{ 0.54}^{ 1.51}$
 &$75.74_{ 0.25}^{ 0.29}$\\
 10    &IRAS\,17038    &$ 0.00_{ 0.00}^{ 0.00}    -  31.29_{ 2.85}^{ 2.25}$    &$ 0.03_{ 0.03}^{ 0.92}    -  31.02_{ 2.54}^
{ 4.33}$    &$21.65_{ 0.91}^{ 0.77}    -   0.00_{ 0.00}^{ 0.00}$    &$ 0.00_{ 0.00}^{ 0.00}    -  16.00_{ 2.13}^{ 1.89}$
 &$81.71_{ 0.35}^{ 0.23}$\\
 11    &IRAS\,17243    &$ 0.49_{ 0.49}^{ 7.03}    -  42.74_{ 2.29}^{ 2.73}$    &$20.13_{ 3.29}^{10.90}    -  14.61_{ 9.18}^
{ 5.25}$    &$17.34_{ 0.95}^{ 0.90}    -   0.00_{ 0.00}^{ 0.00}$    &$ 0.00_{ 0.00}^{ 0.00}    -   4.69_{ 2.60}^{ 1.76}$
 &$83.22_{ 0.29}^{ 0.41}$\\
 12    &IRAS\,19125    &$ 8.22_{ 6.32}^{ 7.72}    -   6.24_{ 4.96}^{ 6.47}$    &$ 8.66_{ 4.28}^{ 8.00}    -  45.53_{20.03}^
{ 5.92}$    &$ 7.98_{ 0.74}^{ 0.98}    -   7.87_{ 1.22}^{ 1.34}$    &$ 0.00_{ 0.00}^{ 0.00}    -  15.50_{ 1.34}^{ 2.65}$
 &$69.72_{ 0.44}^{ 7.00}$\\
 13    &IRAS\,19157    &$ 0.00_{ 0.00}^{ 0.00}    -  63.21_{ 3.12}^{ 5.31}$    &$ 8.43_{ 3.41}^{ 4.28}    -  12.05_{ 8.01}^
{ 6.89}$    &$14.58_{ 1.49}^{ 1.36}    -   0.02_{ 0.02}^{ 0.94}$    &$ 0.00_{ 0.00}^{ 0.00}    -   1.70_{ 1.35}^{ 2.00}$
 &$82.28_{ 0.53}^{ 0.61}$\\
 14    &IRAS\,20056    &$ 0.45_{ 0.45}^{ 2.96}    -  31.83_{ 3.94}^{ 3.62}$    &$34.60_{ 2.66}^{ 2.35}    -   0.30_{ 0.30}^
{ 5.33}$    &$15.86_{ 1.10}^{ 1.04}    -   0.02_{ 0.02}^{ 0.85}$    &$ 0.00_{ 0.00}^{ 0.00}    -  16.93_{ 1.77}^{ 1.80}$
 &$83.48_{ 0.64}^{ 0.20}$\\
 15    &RU\,Cen    &$ 0.00_{ 0.00}^{ 0.00}    -  27.92_{ 2.49}^{ 2.17}$    &$ 1.24_{ 1.06}^{ 2.06}    -  27.81_{ 3.95}^{ 3.
82}$    &$28.63_{ 1.80}^{ 1.83}    -   3.34_{ 2.68}^{ 2.71}$    &$ 0.00_{ 0.00}^{ 0.00}    -  11.06_{ 2.11}^{ 1.71}$    &$8
0.62_{ 0.76}^{ 0.34}$\\
 16    &SAO\,173329    &$ 0.00_{ 0.00}^{ 0.00}    -  52.55_{ 2.35}^{ 1.74}$    &$ 0.02_{ 0.02}^{ 0.50}    -  28.64_{ 2.47}^
{ 2.15}$    &$ 9.21_{ 0.77}^{ 1.09}    -   0.02_{ 0.02}^{ 0.00}$    &$ 0.00_{ 0.00}^{ 0.00}    -   9.56_{ 1.85}^{ 2.63}$
 &$82.54_{ 0.20}^{ 0.44}$\\
 17    &ST\,Pup    &$ 0.00_{ 0.00}^{ 0.00}    -  32.87_{ 1.23}^{ 1.29}$    &$ 0.71_{ 0.66}^{ 1.86}    -  51.93_{ 3.17}^{ 1.
79}$    &$13.19_{ 0.46}^{ 0.57}    -   0.03_{ 0.03}^{ 0.83}$    &$ 0.00_{ 0.00}^{ 0.00}    -   1.27_{ 0.90}^{ 1.16}$    &$5
4.60_{ 0.37}^{ 0.53}$\\
 18    &SU\,Gem    &$ 0.00_{ 0.00}^{ 0.00}    -  58.69_{ 3.77}^{ 3.60}$    &$ 0.70_{ 0.69}^{ 2.88}    -  12.03_{ 5.11}^{ 3.
93}$    &$23.41_{ 1.75}^{ 2.31}    -   2.30_{ 1.96}^{ 3.69}$    &$ 0.00_{ 0.00}^{ 0.00}    -   2.87_{ 2.11}^{ 3.03}$    &$8
8.40_{ 0.43}^{ 0.44}$\\
 19    &SX\,Cen    &$ 0.00_{ 0.00}^{ 0.00}    -  48.37_{ 4.76}^{ 3.30}$    &$ 9.19_{ 5.82}^{11.44}    -  23.70_{23.67}^{12.
70}$    &$10.82_{ 1.97}^{ 3.64}    -   0.53_{ 0.53}^{ 4.29}$    &$ 0.00_{ 0.00}^{ 0.00}    -   7.39_{ 3.15}^{ 4.69}$    &$7
6.16_{ 1.16}^{ 1.50}$\\
 20    &TW\,Cam    &$ 0.00_{ 0.00}^{ 0.00}    -  84.42_{ 2.04}^{ 1.81}$    &$ 0.00_{ 0.00}^{ 0.00}    -   0.00_{ 0.00}^{ 0.
00}$    &$15.20_{ 1.73}^{ 1.47}    -   0.00_{ 0.00}^{ 0.00}$    &$ 0.00_{ 0.00}^{ 0.00}    -   0.37_{ 0.37}^{ 1.84}$    &$9
0.63_{ 0.18}^{ 0.15}$\\
 21    &UY\,CMa    &$ 0.00_{ 0.00}^{ 0.00}    -  15.55_{ 1.96}^{ 2.31}$    &$ 2.54_{ 2.02}^{ 2.53}    -  58.59_{ 3.68}^{ 3.
01}$    &$16.18_{ 1.78}^{ 2.50}    -   4.48_{ 3.42}^{ 2.93}$    &$ 0.00_{ 0.00}^{ 0.00}    -   2.65_{ 1.43}^{ 1.53}$    &$7
8.14_{ 0.42}^{ 0.62}$\\
\hline
\end{tabular}
\end{table}

\begin{table}
\caption{\textit{Erratum for Table\,3 in \citet{gielen09}}: Best-fit parameters deduced from our full spectral fitting.
Listed are the $\chi^2$, dust, and continuum temperatures and their relative fractions. Best-fit parameters deduced from our full spectral fitting. 
The abundances of small and large grains of the various
dust species are given as fractions of the total mass, excluding the dust responsible for the continuum emission.
The last column gives the continuum flux contribution, listed as a percentage of the total integrated flux over the 
full wavelength range.}
\centering
\vspace{0.5cm}
\hspace{0.cm}
\begin{tabular}{lrllllll}
\hline \hline
Name & $\chi^2$ &  $T_{dust1}$ & $T_{dust2}$ & Fraction & $T_{cont1}$ & $T_{cont2}$ & Fraction \\
    &          &     (K)     & (K)         & $T_{dust1}$- $T_{dust2}$    & (K)         & (K)         & $T_{cont1}$-$T_{cont2}$   \\
\hline
EP\,Lyr & 5.4 &$ 100_{50}^{50}$ &$ 200_{50}^{50}$ &$ 0.90_{0.05}^{0.10}- 0.10_{0.10}^{0.05}$ &$ 100_{50}^{50}$ &$ 643_{50}^{302}$ &$ 0.98_{0.04}^{0.01}- 0.02_{0.01}^{0.04}$\\
HD\,52961  & 50.0 &$ 200_{50}^{50}$ &$ 700_{50}^{50}$ &$ 0.90_{0.05}^{0.05}- 0.10_{0.05}^{0.05}$ &$ 100_{50}^{50}$ &$ 1000_{50}^{50}$ &$ 0.99_{0.01}^{0.01}- 0.01_{0.01}^{0.01}$\\
\end{tabular}

\vspace{0.5cm}
\hspace{0.cm}
\begin{tabular}{lccccc}
\hline \hline   
Name & Olivine & Pyroxene & Forsterite & Enstatite & Continuum\\
      & Small  -  Large & Small  -  Large &  Small  -   Large & Small  -   Large &\\
\hline
EP\,Lyr    &$ 0.24_{0.24}^{16.83}    -   8.74_{7.64}^{7.92}$    &$7.17_{4.79}^{13.69}    -  8.09_{7.24}^{12.67}$    &$35.18_{2.78}^{3.04}    -   2.08_{1.89}^{2.61}$    &$ 0.00_{0.00}^{0.00}    -  38.50_{3.46}^{4.30}$    &$57.99_{3.60}^{2.53}$\\

HD\,52961    &$ 0.00_{0.00}^{0.00}   -   0.00_{0.00}^{0.00}$    &$59.17_{0.69}^{0.72}    -  0.00_{0.00}^{0.00}$    &$0.77_{0.69}^{1.46}  -   40.06_{1.62}^{1.02}$    &$ 0.00_{0.00}^{0.00}    -  0.00_{0.00}^{0.00}$    &$68.88_{0.46}^{0.42}$\\

\hline
\end{tabular}
\end{table}

\begin{table}
\caption{\textit{Erratum for Table\,4 in \citet{gielen09b}}:  Best-fit parameters deduced from our full spectral fitting.
Listed are the $\chi^2$, dust, and continuum temperatures and their relative fractions. Best-fit parameters deduced from our full spectral fitting. 
The abundances of small  (0.1\,$\mu$m) and large (2.0\,$\mu$m) grains of the various
dust species are given as fractions of the total mass, excluding the dust responsible for the continuum emission.
The last column gives the continuum flux contribution, listed as a percentage of the total integrated flux over the 
full wavelength range.}
\centering
\vspace{0.5cm}
\hspace{0.cm}
\begin{tabular}{lrllllll}
\hline \hline
Name & $\chi^2$ &  $T_{dust1}$ & $T_{dust2}$ & Fraction & $T_{cont1}$ & $T_{cont2}$ & Fraction \\
    &          &     (K)     & (K)         & $T_{dust1}$- $T_{dust2}$    & (K)         & (K)         & $T_{cont1}$-$T_{cont2}$   \\
\hline
MACHO\,79.5501.13 & 5.1 &$ 200_{50}^{50}$ &$ 725_{50}^{83}$ &$ 0.90_{0.05}^{0.05}- 0.10_{0.05}^{0.05}$ &$ 346_{246}^{184}$ &$ 623_{50}^{99}$ &$ 0.21_{0.18}^{0.69}- 0.79_{0.69}^{0.18}$\\
MACHO\,82.8405.15  & 3.9 &$ 200_{50}^{50}$ &$ 519_{75}^{82}$ &$ 0.90_{0.05 }^{ 0.05}- 0.10_{0.05}^{0.05}$ &$ 300_{50}^{ 50}$ &$ 500_{50}^{50}$ &$ 0.82_{0.02}^{0.03}-0.18_{0.03}^{0.02}$\\
\end{tabular}
\centering
\vspace{0.5cm}
\hspace{0.cm}
\begin{tabular}{lccccc}
\hline \hline   
Name & Olivine & Pyroxene & Forsterite & Enstatite & Continuum\\
      & Small  -  Large & Small  -  Large &  Small  -   Large & Small  -   Large &\\
\hline
MACHO\,79.5501.13    &$0.00_{0.00}^{0.00}    -   0.00_{ 0.00}^{0.00}$    &$48.45_{7.00}^{4.75}    -  0.37_{0.37}^{20.55}$    &$0.00_{0.00}^{0.00}    -   44.54_{3.32}^{3.47}$    &$0.17_{0.17}^{2.86} -  6.47_{4.25}^{5.76}$    &$89.27_{0.86}^{0.70}$\\

MACHO\,82.8405.15    &$ 0.96_{0.95}^{7.13}    -   4.13_{3.97}^{10.91}$    &$52.80_{9.36}^{10.75}    -  4.01_{3.92}^{18.56}$    &$5.50_{2.26}^{2.76} -   20.52_{7.06}^{6.47}$  &$0.14_{0.14}^{3.66} -  11.95_{5.83}^{5.93}$    &$82.63_{1.86}^{1.86}$\\

\hline
\end{tabular}
\end{table}  

\begin{acknowledgements}
CG and HVW acknowledge support of the Fund for Scientific Research of Flanders
(FWO) under the grant G.0178.02. and G.0470.07. This work is based on observations
made with the Spitzer Space Telescope, which is operated by the Jet Propulsion Laboratory,
California Institute of Technology, under a contract with NASA.
\end{acknowledgements}

\bibliographystyle{aa}
\bibliography{/STER/100/cliog/disk28/Artikels/referenties}

\end{document}